\documentstyle[12pt,aaspp4]{article}

\def\ep{\epsilon}

\def\ni{\noindent}

\def\bar{\overline}

\def\OB{\overline{\bf B}}
\def\emf{\overline{\mbox{${\cal E}$}} {}}

\def\emfb{\overline{\mbox{\boldmath ${\cal E}$}} {}}
\def\emfb{\overline{\mbox{\boldmath ${\cal E}$}} {}}

\def\bbE{\overline {\bf E}}
\def\bbh{\overline {\bf h}}

\def\beq{\begin{equation}}
\def\ee{\end{equation}}
\def\lsim{\mathrel{\rlap{\lower4pt\hbox{\hskip1pt$\sim$}}
    \raise1pt\hbox{$<$}}}
\def\gsim{\mathrel{\rlap{\lower4pt\hbox{\hskip1pt$\sim$}}
    \raise1pt\hbox{$>$}}}
\def\bfE{{\bf E}}
\def\bfJ{{\bf J}}

\def\bfA{{\bf A}}
\def\bfa{{\bf a}}
\def\bfe{{\bf e}}
\def\bfB{{\bf B}}
\def\bbJ{\bar {\bf J}}

\def\bB{\overline B}

\def\bV{\overline V}
\def\ts{\times}
\def\lb{\langle}
\def\rb{\rangle}
\def\curl{\nabla {\ts}}

\def\bbV{\bar {\bf V}}

\def\bfv{{\bf v}}

\def\bfV{{\bf V}}
\def\bfj{{\bf j}}

\def\bfe{{\bf e}}

\def\bfb{{\bf b}}

\def\bfB{{\bf B}}

\def\bbB{\overline {\bf B}}
\def\bbA{\overline {\bf A}}

\def\div{\nabla\cdot}

\begin{document}
\title{Toward Coupling Flow Driven and Magnetically Driven Dynamos}
\author{Eric G. Blackman}
\affil{Dept. of Physics and Astronomy, Univ. of Rochester,
    Rochester, NY, 14627, USA}


\begin{abstract}

Most large scale dynamo research for astrophysical rotators 
focuses on  interior 
flow driven helical  dynamos (FDHDs), but larger scale coronal fields most 
directly influence observations.  It is thus important to understand 
the relationship between coronal and interior fields.  
Coronal field relaxation 
is actually a type of magnetically dominated helical 
dynamo (MDHD). MDHDs also occur in fusion plasma devices where they drive a system toward its relaxed state in response to magnetic helicity injection that 
otherwise drives the system away from this state.  
Global scale fields of astrophysical rotators and jets are thus plausibly produced by a direct coupling between an interior FDHD and a coronal MDHD, 
interfaced by magnetic helicity transport through their mutual boundary.  Tracking the magnetic helicity also elucidates how both FDHD and MDHDs evolve and saturate.  
The utility of magnetic helicity is unhampered by its non-gauge
invariance since physical fields can always be recovered.

\end{abstract}


 
\ni {PACS codes: 95.30.Qd;  98.38.Am; 52.55.Ip,
52.30.Cv; 98.35.Eg; 
96.60.Hv}

\section{Introduction}

As emphasized in Ref [\cite{bj06}], from which the present paper 
has evolved, the term ``dynamo'' can induce confusion because it has different meanings not only within the astrophysical context but also when compared to laboratory
plasma dynamos of magnetically dominated environments.  
Many astrophysicists
think of dynamos as the flow driven  amplification of  magnetic energy. 
Others think specifically of the ``large scale'' flow driven 
helical dynamo, where field is amplified on scales larger than the input driving flow.  Researchers immersed in the problem of  flow driven amplification of an initially weak magnetic
field sometimes wonder what role a dynamo could possibly play in a magnetically dominated environment, and yet these are the environments for which laboratory plasma dynamos operate.


In this special issue article I pursue several goals: (1) I
streamline, update and re-organize the conceptual relation between helical dynamos types developed in Ref. [1] 
(2) I strongly emphasize the need to couple FDHDs and MDHDs in astrophysics
culminating with a simple  steady-state relation between the coronal 
and  interior dynamos  to stimulate specific further work.
In section 2, I describe the differences and similarities between
FDHDs and MDHDs, the environments 
in which they occur, and the energy budgets.
In section 3 all of the equations needed to follow their evolution are given.
In sections 4  and 5, I summarize how closed volume and open volume
flow driven helical dynamos (FDHD) 
and magnetic driven helical dynamos (MDHD)
and their respective saturations can be described by different cases of a unified framework that tracks the spatial and spectral
flow of magnetic helicity.  Section 6 outlines the most minimalistic paradigm 
of  how coupled FDHDs and MDHDs operate symbiotically in astrophysics.
Section 7 is the conclusion.

\section{Conceptually Distinguishing Dynamo Types and their Environments}

Three types of dynamos need to be distinguished: (1) Non-helical flow driven (2) FDHD and  (3) MDHD.
Non-helical dynamos [\cite{kazanstev}
-\cite{haugen04}] 
describe the flow-driven amplification of 
magnetic energy via random walk line stretching, folding, and shear.
No helicity of any kind is involved.  The magnetic field is amplified at and below the driving scale, with negligible amplification on larger scales.  I will not discuss  non-helical dynamos further herein.

FDHD describe how an initially weak large scale field is subsequently amplified via strong, large amplitude smaller scale helical velocity fluctuations often in combination with large scale velocity shear
[\cite{moffatt}-
In astrophysics, the source of energy for the velocity flows is ultimately gravity in the case of disks, and gravity + fusion in the case of stars.
This type of dynamo is required to explain the large scale field of the sun and the solar cycle.
An important characteristic of FDHDs is that the large scale magnetic flux is amplified and sustained 
on time and/or spatial scales significantly larger than that of the driving turbulence.  The key quantity which makes
FDHDs work is magnetic field aligned mean electromotive 
force
$\emfb_{||}=(\overline{\bfv\ts\bfb})_{||} \ne 0$, where $||$ indicates along the mean field $\bbB$, and the overbar indicates a spatial, temporal, or ensemble average depending on the specific application.  
This quantity  is the  commonality between FDHD and MDHDs.

An important source of $\alpha$ in the relation 
$\emfb_{||}\propto \alpha \bbB$ for the FDHD
is the kinetic helicity $\overline {\bfv\cdot\curl\bfv}$,  
a pseudoscalar correlation arising, for example,  from the interplay
between stratified turbulence and rotation. 
Inside of a  rotator with an
outward decreasing density, kinetic helicity 
can in principle be sustained as rising eddies expand 
and rotate oppositely to the underlying mean rotation to conserve angular
momentum. Falling eddies rotate in the same direction as the 
mean rotation.  Rising and falling eddies thus
statistically contribute the same sign of 
$\overline{\bfv \cdot \curl \bfv}$ in each hemisphere, with net
opposite signs in each  hemisphere.

As the large scale field in an FDHD grows, a turbulent small scale dynamo
can generate magnetic fluctuations with energy density
comparable to that in the velocity fluctuations before the large
scale field saturates.  The effect of the small scale field growth and boundary terms on the FDHD
are subjects of considerable research.

Unlike FDHDs, MDHDs occur in laboratory plasma fusion confinement configurations
(not to be confused with the laboratory liquid metal experiments designed to test FDHDs
[\cite{gailitisa}
-\cite{sisan}]
). These configurations include Spheromaks and the 
Reversed Field Pinch (RFP) [\cite{bodin90,jiprager}], where the MDHD
  describes the dynamically evolution toward the relaxed state 
by a magnetic field aligned electric field (or equivalently, external injection of one sign of magnetic helicity 
[\cite{jiprager}
-\cite{bellan}]) 
in the presence of external driving away from this state.
 The source of energy  in the laboratory case
is either an electric potential placed across a
gap between electrodes bounding the plasma, which then drives a direct current,
or an externally time dependent current from which 
an electric potential and current generate helical fields 
within the plasma.

Although the magnetic helicity injection 
drives the system away from the relaxed state, 
the driving electric potential injects 
current along an initially toroidal magnetic field, for example, 
in the case of the  
RFP. This generates a poloidal field which becomes unstable to 
 small amplitude fluctuations via 
kink mode instabilities from large currents, or 
tearing modes from current gradients. The fluctuations produce a 
finite $\emfb_{||}$ which reduces the field aligned current
to restore stability by driving a spatial flow of magnetic helicity.
This in turn,  enables the magnetic field structure to
evolve toward the largest helical scale available (subject to boundary
conditions), 
as this would be the lowest energy state [\cite{taylor86}] if it could be reached.  The continuous injection of magnetic helicity 
leads to a quasi-steady dynamical equilibrium, 
with possible sawtooth oscillations. 
as the system is driven away from, and  evolves back toward the relaxed
state. The competing effects are the helicity injection and the MDHD.
If the injection is turned off, the fully  
relaxed state can be reached, but the field
then eventually resistively decays.  Via the  MDHD, the helicity injection 
therefore also sustains the large scale helical field, highlighting that 
that energy is also injected.  Note however that the injected energy 
does not go only into the large scale field but also into small and large scale
velocity flows and heat as well.

Note that classical  Taylor relaxation and fossil field relaxation are often presented as if relaxation is distinct from a dynamo. But these processes are very much linked to the MDHD which provides the dynamical evolution to the relaxed state. The fully relaxed Taylor state is typically found by minimizing the magnetic energy subject to boundary conditions and is achieved only when the driving
is turned off and the dissipation, and velocity flows are neglected.
In general, the relaxed state can include finite non-magnetic
energies as well and generalized MDHD equations can be used to model
the evolution of hydrodynamic and magnetic quantities [e.g.\cite{bf04}].

The coronae above astrophysical rotators 
are likely magnetically dominated 
[\cite{galeev79}
-\cite{fieldrogers93}].
Coronal loops are sites for MDHDs
driven by helical  field injection from the astrophysical rotator below 
[\cite{bf04,blackman05}].  If the helical field injected from below is  produced by a FDHD inside the rotator, the full description of the origin of the large scale fields involves a coupling between the FDHD and MDHD.
The MDHD component provides a direct analogy to the laboratory
case just discussed and is also implicitly revealed 
itself in  specially designed Spheromak experiments 
to study the formation of astrophysical jets [\cite{hsubellan}].
In the lab  direct electric potential drop across the footpoints injects
the helicity. In coronae, this is supplied directly as a buoyant twisted field
rises or by a velocity shear between or within a loop's footpoints that
twists the field (and induces the analogous electric potential to
that of the laboratory case).
The ultimate source of energy for the coronal magnetic fields
are the sources of energy for the FDHD (gravity and fusion as discussed above).

For all extra-terrestrial astrophysical rotators except galaxies,
we observe at most coronal fields, not interior fields.
Coronal holes of the sun are sites of large scale ``open'' 
 along which the solar wind propagates [\cite{sz}].
%
 Jets from accretion engines in 
young stellar objects, active galactic nuclei and 
gamma-ray bursts
or magnetic towers [\cite{pudritz04,lb03,um06}] 
are the analog to coronal holes.
These  large scale coronal fields
can be produced by the MDHD relaxation of smaller scale
loops emerging from a FDHD inside the rotator below [\cite{ws03}].


\section{Unifying Helical Dynamos}

Progress toward understanding helical dynamos and their saturation
has evolved from a combination of numerical
and analytic work that dynamically  incorporates magnetic helicity evolution 
[\cite{ji99},\cite{pfl}-\cite{bs05}].  
Following [\cite{bj06}], 
I review the derivation of the 
evolution equations for mean, fluctuating, and total magnetic helicity respectively,   and show  that  both MDHD
and FDHD  emerge naturally from different limits.

The electric field is
\beq
\bfE=-\nabla\Phi -\partial_t\bfA,
\label{1}
\ee
where $\Phi$ and $\bfA$ are the scalar and vector potentials.
Taking the average (spatial, temporal, or ensemble), and denoting averaged
values by the overbar, we have  
\beq
\bbE=-\nabla{\overline \Phi} -\partial_t\bbA
\label{2}
\ee
Subtracting (\ref{2}) from (\ref{1}) gives the equation
for the fluctuating electric field
\beq
\bfe=-\nabla\phi -\partial_t\bfa
\label{3},
\ee
where $\phi$ and $\bfa$ are the fluctuating scalar and vector potentials.

Using 
$\bfB\cdot \partial_t \bfA= \partial_t(\bfA\cdot \bfB) +\bfE\cdot \bfB -\nabla \cdot (\bfA\ts \bfE)$,
where the latter two terms result 
from Maxwell's equation $\partial_t \bfB=-\curl \bfE$,
and the identity 
$\bfA \cdot \curl \bfE = \bfE\cdot\bfB-\nabla \cdot (\bfA \ts \bfE)$, 
we take the dot product of (\ref{1}) with $\bfB$ to obtain the evolution of the magnetic helicity density
\beq
\partial_t(\bfA\cdot\bfB)= -2(\bfE\cdot\bfB)
-\div(\Phi \bfB + \bfE\ts \bfA)
=-2(\bfE\cdot\bfB)
-\div( 2\Phi \bfB + \bfA \ts \partial_t\bfA).
\label{4}
\ee
Similarly, dotting (\ref{2}) and (\ref{3}) 
with $\bbB$ and $\bfb$ respectively, the time evolution for mean 
large scale and mean fluctuating magnetic helicity densities are
\beq
\partial_t(\bbA\cdot\bbB)= -2\bbE\cdot\bbB
-\div ({\overline\Phi}\ \bbB + \bbE\ts \bbA)
=-2\bbE\cdot\bbB
-\div( 2{\overline \Phi}\  \bbB + \bbA \ts \partial_t\bbA),
\label{5}
\ee
and 
\beq
\partial_t\overline{\bfa\cdot\bfb}= -2\overline{\bfe\cdot\bfb}
-\div(\overline{{\phi} \bfb} + \overline{\bfe\ts \bfa})
=-2\overline{\bfe\cdot\bfb}
-\div (2\overline{ \phi \bfb} + \overline{\bfa\ts \partial_t\bfa}).
\label{6}
\ee

To  eliminate the electric fields 
from (\ref{4}-\ref{6}) we use Ohm's law with only a resistive term: 
\beq
{\bfE}=-\bfV\ts\bfB +\eta \bfJ,
\label{7}
\ee
where $\bfJ$ is the current density and $\eta $ is the resistivity
in appropriate units. 
Taking the average gives 
\beq
{\bbE}=-\emfb -\bbV\ts\bbB+\eta \bbJ,
\label{8}
\ee
where $\emfb\equiv \overline{\bfv\ts\bfb}$ is the turbulent electromotive
force.
Subtracting (\ref{8}) from (\ref{7}) gives  
\beq
{\bfe}=\emfb-\bfv\ts\bfb -\bfv\ts\bbB -\bbV\ts\bfb 
+\eta \bfj.
\label{9}
\ee
Plugging (\ref{7}) into (\ref{4}), gives 
\beq
\partial_t( \bfA\cdot\bfB)= -2\eta (\bfJ\cdot\bfB)
-\div(\Phi \bfB + \bfE\ts \bfA)
=-2\eta (\bfJ\cdot\bfB)
-\div( 2\Phi \bfB + \bfA \ts \partial_t\bfA).
\label{4a}
\ee
Plugging  (\ref{8}) into (\ref{5}) and (\ref{9}) into (\ref{6}) give,
respectively, 
\beq
\partial_t(\bbA\cdot\bbB)= 
2\emfb\cdot\bbB
-2\eta \bbJ\cdot\bbB
-\div({\overline \Phi}\ \bbB 
+ \bbE\ts \bbA)
=2\emfb\cdot\bbB
-2\eta \bbJ\cdot\bbB
-\div( 2{\overline \Phi}\ \bbB + \bbA \ts \partial_t\bbA)
\label{5a}
\ee
and mean fluctuating magnetic helicity
\beq
\partial_t\overline{ \bfa\cdot\bfb}= 
-2\emfb\cdot\bbB-
2\eta\overline{\bfj\cdot\bfb}
-\div(\overline{\phi \bfb} + \overline{\bfe\ts \bfa})
=-2\emfb\cdot\bbB-
2\eta\overline{\bfj\cdot\bfb}
-\div(\overline{2{\phi} \bfb} + \overline{\bfa\ts \partial_t\bfa}).
\label{6a}
\ee
Note that in the absence of flux  and resistive terms,
the growth of the mean large and mean small scale magnetic helicities
are equal and opposite. This will be important for the kinematic regime
of dynamo growth discussed later.

Dotting (\ref{9}) with $\bfb$ and averaging 
reveals the important relation
\beq
{\emfb\cdot\bbB}=\overline{\bfe\cdot\bfb}-\eta\overline{\bfj\cdot\bfb}.
\label{13}
\ee
Both FDHDs and MDHDs can be derived from Eqs. (\ref{5a}-\ref{13}) and
require 
 $\emfb_{||}\equiv{\emfb\cdot\bbB} / \bB^2\ne 0$.

\section{Closed Volume FDHD vs. Closed Volume MDHD }

Here we employ spatial averages and
 distinguish  global closed volume averages, indicated by brackets, 
from local averages (associated with wavenumber $k=1$ in periodic boxes) 
indicated by an overbar.
Global and local averages can be time dependent.  This formalism applies to e.g. the numerical simulations of  Ref. [\cite{b2001}], the analytic study of [\cite{bf02}] for the FDHD, 
and in analytic studies of the MDHD in Refs. [\cite{bf04}] and 
[\cite{blackman05}]. 
For large enough scale separation between fluctuating
and overlined scales,  the distinction between  overline and bracket for
mean small scale scalars and pseudoscalars  can often be ignored.


Globally averaging (\ref{5a}) and (\ref{6a})  gives
\beq
\partial_t\lb\overline{\bfa\cdot\bfb}\rb=
-2\lb\emfb\cdot\bbB\rb
-2\eta\lb\overline{ \bfj\cdot\bfb}\rb
\label{5aa}
\ee
and
\beq
\partial_t\lb\bbA\cdot\bbB\rb=2\lb\emfb\cdot\bbB\rb-2\eta\lb\bbJ\cdot\bbB\rb.
\label{6aa}
\ee

The time evolution equation for $\emfb$ is given by
\beq
\partial_t\emfb=\overline{\partial_t\bfv\ts\bfb} +\overline{\bfv\ts\partial_t\bfb}.
\label{timed}
\ee
and the small scale momentum density and 
induction equations for
$\div\bfv=0$ are 
\begin{equation}
\begin{array}{r}\partial_{t} \bfb = \OB\cdot\nabla\bfv - \bfv\cdot\nabla\OB + 
\curl(\bfv\ts\bfb) 
-\curl\overline{\bfv\ts\bfb} +
\lambda\nabla^{2}\bfb,
\end{array}
\label{c3}
\ee
and
\begin{equation}
\begin{array}{r}
\partial_{t} {v}_q=P_{qi}({\OB}\cdot\nabla{b_i} + {\bfb}\cdot\nabla{\bB_i} 
-\bfv\cdot\nabla v_i+\overline{\bfv\cdot\nabla v_i}
+{\bf b}\cdot\nabla{b}_i-\overline{{\bf b}\cdot\nabla{b_i}})
+ \nu\nabla^{2}{v_q} 
+{f_q},
\end{array}
\label{c2}
\end{equation}
where ${\bf f}$ is a divergence-free 
forcing function,  $\lambda$ is the magnetic diffusivity, 
 $\nu$ is the viscosity, 
and $P_{qi}\equiv (\delta_{qi}-\nabla^{-2}\nabla_q\nabla_i)$
is the projection operator that arises after taking the divergence
of the  momentum density equation to eliminate the 
fluctuating pressure (magnetic + thermal).
For the ``minimal $\tau$'' closure  [\cite{bf02}] (discussed below)
${\bf f}$ can be uncorrelated with $\bfb$ [\cite{sur07a}].
Reynolds rules [\cite{rad}] allow the interchange of averages and time
or spatial derivatives, so the 5th term of 
(\ref{c3}) and the 4th and 6th terms in the parentheses of 
(\ref{c2}) do not contribute when put into correlation averages with a fluctuating quantity.



The manipulations required to obtain a 
practical evolution equation for $\emfb$ by combining
(\ref{timed}-\ref{c2}) 
without using the first order smoothing approximation (FOSA) and
retaining triple correlations using the 
``minimal $\tau$'' closure (MTC),
have been discussed at length elsewhere [e.g. \cite{bf02,bs05}].
The result is
\beq
\partial_t\emf_{||}= {\tilde\alpha}{\bbB^2/|\bbB|}
-{\tilde\beta}{\bbB\cdot\curl\bbB}/|\bbB|-{\tilde \zeta}\emf_{||},
\label{2emf}
\ee
where 
${\tilde\alpha}
=(1/3)(\overline{\bfb\cdot\curl\bfb}-\overline{\bfv\cdot\curl\bfv})$,  ${\tilde\beta} = (1/3)\overline{ \bfv^2}$, and  $\tilde \zeta\sim k_fv$ 
accounts for microphysical dissipation
terms, and triple correlations.
It is often reasonable to 
assume that the left side (\ref{2emf}) vanishes as the $\emfb$ can saturate
long before the dynamo does [\cite{bf02,bb02}]. Then (\ref{2emf}) can be rearranged to give an explicit expression for $\emfb$ similar to what appears
in standard textbooks [\cite{moffatt}], with the  distinction that the time constant $1/{\tilde \zeta}$ here is an eddy turnover time. not 
an approximation to a correlation [\cite{bf02,bf03}].
Although   $\tilde \zeta$ can
be a function of wavenumber 
[\cite{kr96}], even the
simple MTC is 
a significant
improvement over the  traditional 
first order smoothing approximation [\cite{moffatt,parker,vishniac}]
in which triple correlations are ignored. 
The MTC is simpler than the
eddy-damped quasi-normal Markovian closure [\cite{pfl}] 
and has also been successful  for scalar diffusion 
[\cite{bf03},\cite{branfluid}].

The equations to be solved for both the closed FDHD or MDHD 
 are 
(\ref{5aa}), (\ref{6aa}), and 
(\ref{2emf}).  
For the FDHD, kinetic helicity is injected
$\overline{\bfv\cdot\curl\bfv}$ while for the MDHD magnetic helicity or current helicity $\overline{\bfj\cdot\bfb}$ is injected.  
In either case, for a  periodic box, 
the large scale field which grows is fully helical at $k=1$.  
The results and differences between the closed volume
FDHD  and MDHD are summarized  in the next subsections.

\subsection{\it FDHD case} 
The closed box isotropically forced FDHD [\cite{bf02}] 
is an nonlinear version of the $\alpha^2$
dynamo [e.g. \cite{moffatt}] that  incorporates the dynamical backreaction of the
magnetic field on the kinetic helicity driving the flow
and the evolution of magnetic helicity.
Keeping in mind that (\ref{5aa}), (\ref{6aa}), and 
(\ref{2emf}) are the coupled equations to be solved, 
the essence of the dynamo growth and saturation is as follows: the initial
system is driven with a finite 
$\overline{\bfv\cdot\curl\bfv}\simeq
\lb\bfv\cdot\curl\bfv\rb$.
This grows a finite $\emfb_{||}$ which then grows mean large scale 
magnetic helicity of the opposite sign to the driving 
kinetic helicity  via (\ref{6aa}).
At early times, the growth is kinematic because $\tilde{\alpha}$ is dominated
by the kinetic helicity,  $\emfb$ is substantial.
and the resistive terms in (\ref{5aa})
and (\ref{6aa}) are negligible. Thus 
 $\lb\overline{\bfa\cdot\bfb}\rb$ (and 
thus $\lb\overline{\bfj\cdot\bfb}\rb$)  grows  with opposite sign to that associated 
with $\lb\bbA\cdot\bbB\rb$. 
This in turn quenches $\tilde\alpha$ and $\emfb$.
The  helical mean field 
reached at the end of the kinematic phase is estimated by
setting first estimating the mean small scale helical field at quenching
from $|\lb\bfb\cdot\curl\bfb\rb|\sim |\lb\bfv\cdot\curl\bfv\rb|$ and 
then using magnetic helicity conservation to get the large scale 
helical magnetic field. The  is
$\bB^2\sim k_1f_h \lb v^2\rb/k_f$,
where $f_h\equiv |\lb\bfv\cdot\curl\bfv\rb|/\lb v^2\rb$, and
$k_1$ is the wave number associated with $\bB$. 

Because the resistive term in (\ref{5aa}) is larger than that
of (\ref{6aa}) in the kinematic phase, 
$|\lb\bbA\cdot\bbB\rb|$ can grow to exceed $|\lb\overline{\bfa\cdot\bfb}\rb|$
beyond the kinematic regime, but at a microphysical resistively limited rate.
Eventually however, the system reaches a true steady state in which growth
balances decay. This occurs 
when the value of the mean current helicity grows large enough such that 
the dissipation terms in (\ref{5aa}) and (\ref{6aa}) are equal.

For $\alpha^2$ dynamos of  standard texts [\cite{moffatt}],
the equation for the mean magnetic field is solved with an
imposed $\emfb$ and linear growth
results. Ref. [\cite{bb03}] emphasizes that 
magnetic helicity is not conserved even in the kinematic regime 
(neither in the equations nor in the diagrams)
for dynamos in the standard texts. In the modern version just discussed, 
the additional time dependent equations for
mean small scale helicity evolution $\emfb$
 evolution are coupled into the theory dynamically and magnetic helicity
evolution is properly evolved.
The approach can be generalized to an $\alpha-\Omega$ dynamo [\cite{bb02}]
where the mean large scale magnetic helicity evolution
equation is be replaced by the vector equation for $\bbB$ 
[\cite{bb02}].

The two-scale analytic approach has been generalized to a four 
scale approach [\cite{blackman03}] to assess whether the mean small scale 
magnetic helicity is more dominated by the dissipation scale or 
 forcing scale (with the mean large scale magnetic helicity 
migrating toward the even larger box scale.)
The analysis shows that the mean small scale magnetic helicity is first
dominated near the resistive scale but migrates toward the forcing
scale before the end of the kinematic regime.
Numerical simulations of helical dynamos
in a periodic box [\cite{b2001,maronblackman}]
also show that the magnetic helicity in saturation
peaks with opposite signs  at the forcing scale and
box scale respectively. In short, 
the system achieves bihelical  equilibrium  in which
the mean small and mean large scale magnetic helicities are dominated by  
the largest scales available to them respectively. The driving
kinetic helicity ensures that these two scales are distinct
and prevents the mean small scale magnetic helicity from migrating 
to the box scale.

\subsection{\it MDHD case}

For the MDHD case, mean small scale current helicity 
$\lb\overline{\bfj\cdot\bfb}\rb=k_f^2\lb\overline{\bfa\cdot\bfb}\rb$
is injected into the $\tilde\alpha$ of 
$\emfb$ in (\ref{2emf}).
Again, because  (\ref{5aa}), (\ref{6aa}), and 
(\ref{2emf}) are the equations to be solved.  
like the FDHD, the MDHD growth of mean large scale and mean small magnetic helicity
is also mediated by the difference between the mean small scale kinetic and current helicities. 

Like the FDHD, for the MDHD
 the mean large scale and mean small scale magnetic helicities have growth rates of 
opposite sign. But for the MDHD,  the mean large
scale magnetic helicity grows with the SAME sign as that of the injected
helicity, not the opposite sign as in the FDHD case.
That the MDHD drives magnetic helicity to large scales exemplifies
that the MDHD is essentially a generalized dynamical Taylor relaxation. 
It is thus appropriate to refer to the MDHD  as ``dynamical magnetic relaxation.''
The lowest energy state of a unihelical configuration is one in which the magnetic helicity resides at the largest scale, subject to available boundary conditions or imposed forcing away from the relaxed state.
In this context,  both the steadily forced case and the case
 for which forcing is turned off have been studied [\cite{bf04,blackman05}]. 

In the forced case, the large scale magnetic helicity grows to about 1/2 of the forced 
value of small scale  magnetic helicity 
kinematically (independent of dissipation) after which the true  steady state evolves via  a {\it viscously} limited phase. This viscously limited  
phase is analogous to the resistively limited phase of the FDHD and arises
because the kinetic helicity is the backreactor in the MDHD case and so its
dissipation facilitates the slow growth phases after the kinematic regime.
Eventually a steady MDHD results, analogous to the FDHD case. In the MDHD case like the FDHD
case most of the magnetic energy resides on the largest $k=1$ scale in the saturated steady state. Unlike the FDHD case, the forcing scale and the $k=1$ scale magnetic helicities have the same sign. In addition, the total magnetic energy at the forcing scale dominates the kinetic energy there.
The saturation to the steady state in the forced case arises 
 because the injected magnetic helicity also
induces a growth of kinetic helicity at the injection scale and this
which plays the role of the backreacting agent.
This depletes the $\tilde \alpha$ in (\ref{2emf})
and saturates the growth of the large scale magnetic helicity.
The roles of the kinetic and current helicities are therefore 
reversed for the MDHD compared to the FDHD.

For the case in which the small scale helicity is injected only initially,
the kinematic phase proceeds similar to the forced phase, but then 
all quantities eventually decay, with the large scale helicity decaying the slowest.



The MDHD regime, unlike the FDHD regime, has not been
fully  tested  with  3-D MHD numerical experiments,
but similar simulations in  a periodic box 
for which  magnetic helicity is injected at some 
wavenumber, say $k_f\sim 5$, with an initially negligible velocity
would be appropriate.
The overall evolution of the magnetic helicity and kinetic helicity
spectra could then be measured as a function of time.

\section{Open Volume FDHD vs. Open Volume MDHD}

Here we assume overlined averages are taken  over 
large scales with respect to fluctuating quantities 
but over scales equal to or smaller than the scale of the bracketed averages.
For example, in an accretion disk,  
the bracketed averages could be taken over an entire hemisphere
within the disk, 
whereas the overlined averages could be taken over full azimuth and
half scale height, but remain local in radius.
Note that neither average would include the corona, and so global
vertical fluxes can remain.

In the limit
that time evolution and resistive terms
are ignored but the divergence
terms are kept, Eqs. (\ref{5a}) and 
(\ref{6a}) give
\beq
0=2\lb\emfb\cdot\bbB\rb
-\div\lb{\overline\Phi}\  \bbB + \bbE\ts \bbA\rb
\label{26}
\ee
and 
\beq
0=
-2\lb\emfb\cdot\bbB\rb
-\div\lb\overline{\phi \bfb} + \overline{\bfe\ts \bfa}\rb.
\label{27}
\ee
Combining these two  equations reveals that  the fluxes
of mean large and mean small scale helicity through the system
boundary are equal and opposite. This is important for
 an FDHD inside of an astrophysical rotator,  an MDHD outside of a
astrophysical rotator, and an MDHD  in laboratory
plasmas: The  $\emfb_{||}$ for  these cases is  supplied 
by a helicity flux.

\subsection{\it Open FDHD case}

If helical motions  sustain  kinetic helicity inside of the rotator, then 
 $\emfb$ is sustained by $\tilde \alpha$
from (\ref{2emf}), and the averaging in (\ref{26}) and (\ref{27})
is taken over scales less than or equal to the interior  hemisphere of the rotator.
A steady-state with open boundaries would imply that 
that the rotator supplies each hemisphere of 
its corona with one sign of magnetic helicity on 
large scales and the other on smaller scales 
[\cite{bb03},\cite{bf00b}] . 
 The coronal energy deposition rate 
associated with these helicity fluxes [\cite{bf00b}]
is consistent   time-averaged steady coronae of the sun [\cite{sz}] 
and AGN accretion disks 
[\cite{hm93,fieldrogers93}].
The bihelical nature of the field
and the sign dependence of the injected helicity for any single 
structure can also be influenced  by  whether additional localized 
surface shear operates on a scale larger or smaller than that of
a given loop's footpoint separation [\cite{demoulin02}].
A statistical approach is therefore needed to infer the dominant scale dependent sign 
of  magnetic  helicity.

\subsection{\it Open MDHD case}

In the context of an astrophysical rotator, the MDHD
case applies when the volume averages of 
(\ref{26}) and (\ref{27}) are  that  of a corona.
The FDHD supplies magnetic helicity via the flux terms
which then drive the coronal MDHD.
The evolution of coronal magnetic structures
 is analogous to the evolution of magnetically dominated laboratory plasma configurations 
[\cite{bellan,hsubellan}] subject to injection of magnetic helicity.
The experiment of Ref. [\cite{hsubellan}] 
reveals a direct analogy to helical loops of flux rising
into an astrophysical corona from its rotator below.
The loops coalesce at the symmetry axis and form
a magnetic tower. For large enough helicity injection, 
the tower can break off a Spheromak blob from the kink instability. 
The helicity flux from the loops  seeds  
subsequent dynamical magnetic relaxation  via am MDHD. 
The relaxation opens field lines that form coronal holes or jets.

The astrophysical corona, taken as a single entity, 
can also be modeled as a statistical aggregate 
of  loops and the corona can be thought of as a single dynamical entity [\cite{bf04}]
which statistically receives 
 injected helicity of both signs in each hemisphere 
such that 
bihelical relaxation [\cite{blackman05}]
applies.  However, for a single  structure within that corona,
the net sign may be positive or negative. It is in this sense 
that the guiding principles of the MDHD in a corona for each structure can then 
be understood by analogy to laboratory plasma configurations.
The most direct analogy
comes from comparing a toroidal laboratory configuration such as an RFP
with a single coronal loop of magnetic flux, subject to footpoint twisting that injects helicity into the loop.  We can think of each loop as a cut torus with
a net injection of helicity of one sign.

It is useful to  discuss  the MDHD in the context of a torus, streamlining
the treatment of [\cite{bj06}].
 Overlined mean quantities are time averages
and spatial averages over  periodic directions $\phi$ (locally $\hat{\bf z})$ and $\theta$,
but not over  radius $r$ (where $r=0$  corresponds to an azimuthal ring 
at the center of the torus' cross section.) 
The steady-state limit of (\ref{6a}) is
\beq
\emfb_{||}=
{\bbB\over \bB^2}(\nabla\cdot\bbh-\eta\overline{\bfj\cdot\bfb}),
\label{14b}
\ee
where $\bbh\equiv - (\overline{
\phi \bfb}+{1\over 2}\overline{\bfa \ts\partial_t\bfa})$.
Dotting (\ref{14b}) with $\bbB$, and using (\ref{8})
gives
\beq
\emfb\cdot \bbB = 
\div {\bbh}-\eta\overline{\bfj\cdot\bfb}  
=\eta \bbJ\cdot\bbB - \bbE\cdot\bbB,
\label{new8} 
\ee
where, from Eqn. (\ref{5}), we also have 
\beq
\bbE\cdot\bbB=
-\div( {\overline \Phi}\ \bbB + {1\over2}\bbA \ts \partial_t\bbA)
=
-\div( {\overline \Phi}\ \bbB),
\label{new8bb}
\ee
where the latter equality  follows for the assumed steady-state.

The RFP dynamo 
emerges when an 
$\bbE$ of sufficient strength 
is externally applied along the initial toroidal magnetic field.
A finite $\emfb_{||}$ then results from fluctuations
induced by tearing or kink mode instabilities.
Were there no  $\emfb$,  the two terms on the right of 
(\ref{new8}) would balance.
For sufficiently large applied $\bbE_{||}$, 
RFP experiments   [\cite{jiprager,bodin90,caramana84,japs94}]
reveal that $\bbE\cdot\bbB=\eta \bbJ\cdot\bbB > 0$ only at a single radius
$0<r=r_c<a$, 
where $a$  is  the minor radius of the torus and $r_c$ is measured from the
toroidal axis. 
For $r< r_c$, $\bbE\cdot\bbB > \eta \bbJ\cdot \bbB > 0$ and for $r> r_c,\ \ $ 
$ \eta \bbJ\cdot\bbB > 0 >\bbE\cdot\bbB$.  Excluding pressure gradient and inertial terms in Ohm's law, 
such measurements imply that $\emfb_{||}\ne 0$. 
Since $\eta\bbJ\cdot\bbB-\bbE\cdot\bbB$
changes sign from negative to positive moving outward through $r_c$
(while $\bbJ\cdot\bbB$ keeps the same sign), 
$\emfb_{||}$ must 
also change  from negative to positive across $r=r_c$.
If the third term in (\ref{14b}) is negligible, 
(\ref{14b}) shows that $\div {\bbh}$ must change sign through $r_c$.

Volume integrating (\ref{new8})  eliminates the $r$ dependence.
Using (\ref{new8bb}), this
gives
\beq
\int \emfb\cdot \bbB dV= \int{\bbh}\cdot d{\bf S}
=\int(\eta\bbJ\cdot\bbB - \bbE\cdot\bbB)dV
=\int\eta\bbJ\cdot\bbB dV + 
\int 
( {\overline \Phi}\ \bbB
)\cdot d{\bf S},
\label{new8b}
\ee
dropping the third term of (\ref{14b}) 
(justified by experiment [\cite{jiprager,bodin90,caramana84,japs94}]]),  
and using Gauss' theorem to obtain surface integrals, 
keeping in mind that for doubly connected topologies 
this requires $\bbh$ to be  analytic everywhere. The latter is  ensured
because our averaged quantities depend only on radius.
Taking the surface integral in (\ref{new8b})  over the cut faces of the torus
we have 
\beq
 \int{\overline \Phi} \ \bbB \cdot d{\bf S}
=\Delta {\overline \Phi}
\int{\overline \bfB}\cdot d{\bf S}
=V_s\Psi_s,
\label{voltage}
\ee
where we have assumed that $\overline \Phi$ has a constant value
on each of the surface faces, $V_s=\Delta {\overline\Phi}=\int
\bbE\cdot d{\bf z}$ is the externally applied
potential drop between  toroidal faces,  
and $\Psi_s$ is the toroidal magnetic flux through both surfaces. 
Eq. (\ref{voltage}) represents the helicity
injection; a similar term drives the instabilities of Spheromak
experiments of Ref. [\cite{hsubellan}].

The parallel component of the electromotive force can  be written $\emfb_{||} =\alpha\bbB$.
The pseudoscalar $\alpha$ for the laboratory case is given from (\ref{13}) and
(\ref{new8}) by
\beq
\alpha = 
{\emfb\cdot \bbB\over \bB^2}\sim 
{1\over \bB^2}\div\bbh=
{\overline{\bfe\cdot \bfb}\over \bB^2}\simeq 
{\overline{\bfe_\perp\cdot \bfb_\perp}\over \bB^2},
\label{20}
\ee 
where the last similarity follows because the fluctuations
are primarily perpendicular to the strong mean fields. 
The  measured right side of (\ref{20}) is consistent with the 
$\emfb$ needed for MDHD models
[\cite{japs94}].


\subsection{\it Time-dependent, Open, FDHD}

In general, both the time dependent and the  flux terms 
in equation (\ref{5a}) and (\ref{6a}) can be included dynamically
rather than taking the simpler steady-state of the previous sections.
Two recent calculations of FDHDs, applied to the interiors of
astrophysical disks, incorporate the time dependence using different sets of approximations.

In the context of the Galaxy, Refs. [\cite{sur07a}] and [\cite{shukurov06}] 
solved the mean field induction equation for $\bbB$ with
$\emfb_{||}$ determined from setting 
$\partial_t \emfb_{||}=0$ in (\ref{2}).
The $\emfb_{||}$ involves the difference between the kinetic and current 
helicities which can be related to
small scale magnetic helicity in the Coulomb gauge.
Refs. [\cite{sur07a}] and [\cite{shukurov06}] 
 formally use a gauge invariant
helicity density [\cite{sb06}] $\chi$  
to replace the magnetic helicity density but
the role of the boundary terms is conceptually independent of this.  
Effectively, the induction  equation is solved for $\bbB$ 
(which depends on $\emfb_{||}$ and thus $\lb\overline{\bfa\cdot\bfb}\rb$)
and Eq. (\ref{6a}) for $\lb\overline{\bfa\cdot\bfb}\rb$.
The divergence term in (\ref{6a}) can be rigorously [\cite{sb06}]
replaced with one of the form 
$\propto \nabla\cdot (\chi \bbV)$, where $\bbV = (0,0,\bV_z) $ is the mean
velocity advecting the small scale helicity out of the volume.
This mean velocity also appears in the induction equation for $\bbB$, 
highlighting that the  loss terms in the small scale helicity equation also
implies advective loss of mean field. For the 
estimates in Refs. [\cite{sur07a}] and [\cite{shukurov06}] 
the distinction between $\chi$ and $\lb\bfa\cdot\bfb \rb$ in the Coulomb
gauge is not essential.

This approach supports the concept
[\cite{bf00a}] 
that a flow of small scale helicity toward the boundary may help
to alleviate the backreaction 
of the small  scale  magnetic helicity on the 
kinetic helicity which drives the dynamo in  $\emfb_{||}$. 
However, if $\bV_z$ is too large, it may
carry away too much of the  mean field which the dynamo is trying to grow
in the first place.
In general, more work is needed to calculate $\bV_z$ from
first principles, and its effect on large and small scale fields.
The helicity flows in the time-dependent case can also seed a
time dependent coronal MDHD.

A more restrictive time dependent FDHD dynamo that includes boundary terms, maintains the time dependence in (\ref{5a}), but implicitly
assumes that equation (\ref{6a}) reaches a steady-state has also been
studied [\cite{vishniac}].  
Although the approach involves assumptions that have 
now been avoided in more general calculations of helicity fluxes 
[\cite{sb04}] 
(one being  the first order smoothing approximation which can be avoided by the MTC 
discussed earlier), Ref.[\cite{vishniac}]
does identify how a time dependent flow-driven dynamo 
in a Keplerian shear flow might be sustained by a  magnetic helicity flux.
Here the relevant forms of 
(\ref{5a}) and (\ref{6a}) are
\beq
\partial_t (\bbA\cdot\bbB)= 
2\emfb\cdot\bbB
-\div({\overline\Phi}\ \bbB + \bbE\ts \bbA)\
\label{26a}
\ee
and 
\beq
0=
-2\emfb\cdot\bbB
-\div(\overline{{\phi} \bfb} + \overline{\bfe\ts \bfa}).
\label{27a}
\ee
Using (\ref{27a}),  $\emfb_{||}$ 
can be directly written in terms of the  
small scale helicity flux as 
\beq
\emfb_{||}=-{\bbB\over B^2}
\nabla\cdot(\overline{{\phi} \bfb} + \overline{\bfe\ts \bfa})
=
-{\bbB\over B^2}
\nabla\cdot \overline{(-\nabla\phi +\bfe)\ts\bfa}.
\label{vc1}
\ee
then use
(\ref{vc1})  
in the equation for the mean magnetic field applied to an accretion disk
whose mean quantities are axisymmetric. 
The mean field equation is 
\beq
\partial_t\bbB=\curl \emfb +\curl (\bbV\ts\bbB)+\lambda \nabla^2\bbB.
\label{vc2}
\ee
Solving (\ref{vc2}) requires use of (\ref{vc1}).  
Ref. [\cite{vishniac}] invokes a correlation time $\tau_c$ such that  $\bfa \simeq -(\bfe +\nabla\phi)\tau_c$, 
(note: Ref. [\cite{vishniac}] defines
 $\bfe_{mf}\equiv -\bfe$ and work with $\bfe_{mf}$).  
This reduces the last term of (\ref{vc1}) to 
\beq
2{\bbB\over B^2}\tau_c
\nabla\cdot \overline{\bfe\ts\nabla\phi}\equiv -{\bbB\over B^2}\nabla \cdot {\bf J}_H,
\label{helcur}
\ee
which defines the helicity flux $\bfJ_H$. (Ref. [\cite{vishniac}] is 
missing a factor of 2).
Assuming incompressible flow for the fluctuations, 
\beq
J_{H,i}\sim 2l^2 \tau_c \overline{\bbB\cdot\nabla v_i\bbB\cdot {\curl \bfv}},
\label{34}
\ee
where $l$ is the outer turbulent scale.
This current  is then used in (\ref{vc2}) to allow growth
of $\bbB$.

 Refs. [\cite{sb04,bs05}]
 show that (\ref{34})  is one of a number
of current terms that emerge in a more general calculation which avoids
the first order smoothing approximation (see section 3.2) 
with additional fluxes arising when $\bbV$ is included in $\bfe$.
Nevertheless, the importance of  Eq. (\ref{34}) 
is that it shows how dynamo growth of  the large scale field can be driven entirely by the small scale helicity flux without any
kinetic helicity.  Determining whether this
works in practice is an active area of research.
Refs. [\cite{bsan}], [\cite{b2005}], and [\cite{sur07b}] collectively 
show that the  Vishniac-Cho flux  can sustain large scale field growth
 via boundary flow 
in the absence of kinetic helicity, 
 but requires the field to already
exceed  a double-digit percentage
of the equipartition value of the turbulent velocity before this effect kicks 
in.

The role of the boundary flux to alleviate catastrophic dynamo quenching
[\cite{bf00a}] may actually be 
more  important for the Sun than the
Galaxy since the former requires 
a rapid, unquenched, cycle period [\cite{bb03}].  
However, as also emphasized in the previous subsections, 
large scale helicity flux likely accompanies any small scale helicity
flux. Significant loss of the large scale field would imply
removal of the large scale field that the dynamo is invoked
to generate inside the rotator thereby lowering its maximum value
inside the rotator. Care in identifying the relative amount of 
large and small scale helicity flux is warranted.

\section{Coupling Flow Driven and Magnetically Driven Dynamos}

As emphasized earlier, 
global scale field growth in astrophysical rotators plausibly involves
a coupling of the interior FDHD to the exterior dynamo MDHD. The fields which are large scale with respect to the FDHD in the interior
are small scale with respect to the corona and the large 
scale fields of coronae are on a larger ``global'' scale.
An example of an FDHD produced large scale  
field  would be  a flux tube with each footpoint
on the scale of order the thickness of the turbulent zone of the underlying rotator, i.e. a disk scale height, or convection zone thickness.

In a steady state, both signs of magnetic helicity would be injected into the corona on separate scales, as per (\ref{26}) and (\ref{27}).
Consider here an ensemble of individual structures 
whose  mean ``large scale''   
helicity injected from the interior to the corona is coherent
in order to further exploit the  analogy to the steady-state laboratory MDHD case discussed in section 5.
We  take
the divergence term in (\ref{26}) to indicate the flux of mean 
``small scale'' helicity  to the corona. That is, we posit a steady state with the correspondence
\beq
 \div\lb{\overline\Phi}\ \bbB + \bbE\ts \bbA\rb_{int}=
-\div\lb{\phi} \bfb + \bfe\ts \bfa\rb_{cor}.
\label{34a}
\ee
Then 
\beq
0=2\lb\emfb\cdot\bbB\rb_{cor}
-\div\lb{\overline\Phi}\ \bbB + \bbE\ts \bbA\rb_{cor}
\label{35}
\ee
and 
\beq
0=
-2\lb\emfb\cdot\bbB\rb_{cor}
-\div\lb{\phi} \bfb + \bfe\ts \bfa\rb_{cor}.
\label{36}
\ee
The subscripts ``int'' and ``cor'' indicate  averaging scales chosen
to be taken within  a hemisphere interior to the rotator and in the corona respectively. The mean scales for the corona structure and the disk structure
can be quite different. When both the disk and the
coronal region are assumed to be in a mutually steady state,  
Eqs. (\ref{34a}-\ref{36}) and (\ref{26a}-\ref{27a}) imply
\beq
\lb\emfb\cdot\bbB\rb_{cor}=\lb\emfb\cdot\bbB\rb_{rot}.
\label{37}
\ee
The coronal electromotive force is driven by the flux into the corona
which itself can be ultimately driven by the helical turbulence within the rotator.
Since it is $\lb\emfb\cdot\bbB\rb_{rot}$ which sources the large scale field of the corona, the assumed steady state implies a constant generation and loss of large scale magnetic helicity in the corona. 
The quasi-steady state can be  established for a fixed coronal volume via loss 
in e.g. a Poynting flux jet.

The amount of helicity injected into the corona also 
provides a lower limit on the energy injected into astrophysical coronae
and is consistent with that needed for the sun and active galactic nuclei
[\cite{bf00b}].  The role of CMEs for the sun deserves special
attention as a mechanism by which magnetic helicity is ejected into the corona.
Ref. [\cite{bb03}] suggests that both signs of magnetic helicity
are ejected by CMEs, in both hemispheres. This serves to keep the 
solar cycle fast, but also lowers the net saturation value of the mean field in the
solar interior that would otherwise arise without ejection. 
The MDHD is the process by which CMEs relax in the corona.

\section{Conclusions}

By following  the dynamical evolution of magnetic helicity, a
unifying  framework  for flow driven and magnetically driven 
helical dynamos and their nonlinear saturation emerges.  
Both FDHDs and MDHDs fit into this framework.  Standard $\alpha^2$ and 
$\alpha-\Omega$ type dynamos are  FDHDs  while
coronal dynamical magnetic relaxation (or dynamical Taylor relaxation), 
and laboratory plasma dynamos are MDHDs.  

 A unifying principle for FDHD and MDHD  is that both require a turbulent electromotive force aligned with the mean magnetic field.
For closed FDHD fed by kinetic helicity, the kinetic helicity 
drives large scale magnetic helicity growth which in turn 
corresponds to the large scale magnetic field growth. But since no net
magnetic helicity is injected into the system, the opposite sign of magnetic helicity must grow on small scales to compensate. For the closed MDHD, the system
is injected with magnetic helicity of one sign on small scales 
and the dynamo proceeds by relaxing this field to large scales whilst
depleting the helicity from small scales.  

In astrophysics,  open FDHDs and open MDHDs are coupled
via the boundary between an astrophysical rotator and its corona.
The FDHD can feed the MDHD 
via magnetic  helicity injection at coronal loop footpoints.
Although isolated FDHDs have received most of the attention in astrophysics,
it is in fact
coronal relaxation via the MDHD that produces the fields most directly observed.
More work is definitely needed to understand how these principles apply in specific 
systems.

Finally, one might raise the concern that  
because magnetic helicity is a gauge non-invariant quantity, 
its prominent conceptual role in recent dynamo theory work is puzzling.
In fact, magnetic helicity evolution can be employed
entirely as a calculational tool with a  convenient  gauge chosen.
At the end of the day, one can always convert back to the magnetic and 
electric fields. 
It is only when one specifically demands a 
physically measurable variant of magnetic helicity [\cite{berger,finn}]
for its own sake, that the issue of defining  a gauge invariant quantity arises.

{\bf Acknowledgments}:
EGB acknowledges support from 
NSF grants AFT-0406799, AST-0406823 and NASA grant ATP04-0000-0016.


\enumerate

\bibitem{bj06}  E.G.  Blackman  \& H. Ji, 
MNRAS, {\bf 369}, 1837 (2006)

\bibitem
{kazanstev} A.P. Kazanstev, 
Sov. Phys. JETP,  {\bf 26} 1031 (1968)

\bibitem{maroncowley}
J. Maron, S. Cowley, J. McWilliams, Astrophys. J., 
{\bf 603} 569 (2004)

\bibitem{schek02}A.A. Schekochihin, S.C.  Cowley, G.W. Hammett, 
G.W., J.L. {Maron , J.C. McWilliams,  New J. Phys., {\bf 4}, 84 (2002)

\bibitem{schek04} A.A. Schekochihin, 
S.C. Cowley, S.F. Taylor, J.L. Maron, \& J.C. McWilliams 
ApJ {\bf 612}, 276 (2004)

\bibitem{haugen03} 
N.E.L. Haugen, 
A. Brandenburg, W. Dobler,   Astrophys. J. Lett., {\bf 597}, L141 (2003)
 
\bibitem{haugen04} 
 N.E.L. Haugen, 
A. Brandenburg, 
Phys. Rev. E., {\bf 70}, 036408 (2004)

\bibitem{moffatt} H.K. Moffatt, {\sl Magnetic
Field Generation in Electrically Conducting Fluids}, 
(Cambridge University Press, Cambridge, 1978)

\bibitem{parker}  
E.N. Parker, {\it Cosmical Magnetic Fields}, (Oxford: Clarendon
Press, 1979)

\bibitem{krause}   F. Krause \& K.-H. R\"adler, 
{\it Mean-field Magnetohydrodynamics and Dynamo Theory}, 
(Pergamon Press, New York, 1980)

\bibitem{zeldovich83} 
Ya. B. Zeldovich , A.A. Ruzmaikin, \& D.D. Sokoloff, {\sl Magnetic Fields in Astrophysics}, 
(Gordon and Breach, New York, 1983)

\bibitem
{gailitisa} A. Gailitis, 
O. Lielausis, E. Platacis, G. Gerbeth,  \& F. Stefani,   Reviews of 
Modern Physics, {\bf 74}, 973 (2002)

\bibitem
{gailitisb}A. Gailitis, 
O. Lielausis, E. Platacis, G. Gerbeth, \& F. Stefani, Surveys in 
Geophysics, {\bf 24}, 247 (2003)

\bibitem
{jimri}H. Ji, J. Goodman\& 
A. Kageyama, \mnras, {\bf 325}, L1 (2001)

\bibitem
{noguchi} K. Noguchi, V.I.Pariev, 
S.A. Colgate, H.F. Beckley, \& J. Nordhaus,  ApJ, {\bf 575}, 
1151 (2002)

\bibitem
{peffley} N.L. Peffley, A.G.Goumilevski, A.B. Cawthrone,  \& D.P. 
Lathrop  Geophysical 
Journal International, {\bf 142}, 52 (2000)

\bibitem
{sisan} D.R. Sisan, et al.
PRL, {\bf 93}, 114502 (2004)

\bibitem{bodin90} H.A.B. Bodin and A.A. Newton, Nuclear Fusion
{\bf 20}, 1717 (1990).

\bibitem{jiprager} 
 H. Ji \& S.C. Prager,  {Magnetohydrodynamics} {\bf 38}, 191  (2002)

\bibitem{ji99} H. Ji, Physical Review 
Letters, {\bf 83} 3198, (1999)

\bibitem{ortolani93} 
S. Ortolani  \& D.D. Schnack, 
{\it Magnetohydrodynamics of Plasma Relaxation}
(World Scientific: Singapore, 1993)

\bibitem{strauss85} 
H.R. Strauss, Phys. Fluids, {\bf 28}, 2786 (1985)

\bibitem{strauss86} 
H.R. Strauss, Phys. Fluids, {\bf 29}, 3008 (1986)

\bibitem
{bhattacharjee86} 
A. Bhattacharjee \& E. Hameiri  Phys. Rev. Lett. 
{\bf 57}, 206 (1986)

\bibitem
{holmesetal88} 
J.A. Holmes, B.A. Carreras, P.H. Diamond, \& V.E. Lynch, 
Phys. Fluids, {\bf 31}, 1166  (1988)

\bibitem{gd1}  
A.V. Gruzinov \& P.H. Diamond, Phys. Plasmas, {\bf 2}, 1941 (1995)

\bibitem{by}  A. Bhattacharjee \& Y. Yuan, Astrophys. J.,  
{\bf 449}, 739 (1995)

\bibitem{bellan} P.M. Bellan, {\sl Spheromaks}, 
(Imperial College Press, London, 2000)

\bibitem{taylor86}
J.B. Taylor, Rev. Mod. Phys., {\bf 58}, 741 (1986) 

\bibitem{galeev79} 
A.A. Galeev, R. Rosner, G.S. Vaiana, Astrophys. J., {\bf 229} 318 (1979)

\bibitem{sz}
C.J. Schrijver  \& C. Zwaan, {\it Solar and Stellar Magnetic Activity},
(Cambridge: Cambridge Univ. Press, 2000)

\bibitem{hm93} F. Haardt \& 
L. Maraschi, ApJ {\bf 413} 507,  (1993)

\bibitem{fieldrogers93} 
G.B. Field \& R.D. Rogers, R.D. Astrophys. J., {\bf 403} 94 (1993) 

\bibitem{bf04} E.G. Blackman \& 
G.B. Field, Physics of Plasmas, {\bf 11}, 3264 (2004)
 
\bibitem{blackman05}E.G.  Blackman,
Physics of Plasmas, {\bf 12}, 2304 (2005)

\bibitem
{hsubellan} S.C. Hsu \& P.M. Bellan,  MNRAS, {\bf 334}, 257 (2002)

\bibitem{pudritz04} R.E. Pudritz,
Les Houches Summer School {\bf 78}
187 (2004)

\bibitem{lb03} D. Lynden-Bell,  
MNRAS, {\bf 341} 1360 (2003)

\bibitem
{um06} D.A. Uzdensky 
\& A.I. MacFadyen,  ApJ, {\bf 647} 1192 (2006)

\bibitem
{ws03} Y.-M. Wang \& 
N.R. Sheeley, ApJ, {\bf 599}, 1404 (2003)

\bibitem{pfl}
A. Pouquet, U. Frisch,  J. L\'eorat,   J. Fluid Mech., {\bf 77}  321 (1976)

\bibitem{kleeorin82}
N.I. Kleeorin  \& A.A. Ruzmaikin, 
{Magnetohydrodynamics}, {\bf 2}, {17}, (1982)

\bibitem
{fieldblackman} 
G.B. Field  \&  E.G. Blackman, Ap. J., { \bf 572} 685 (2002)

\bibitem{kr}
N. Kleeorin \& I. Rogachevskii,  Phys. Rev. E., {\bf 59}, 6724 (1999)

\bibitem{bf00a} E.G. Blackman, 
G.B. Field, ApJ,  {\bf 534}, 984 (2000)

\bibitem{rk} I. Rogachevskii 
\& N. Kleeorin, Phys. Rev. E., {\bf 64}, 56307 (2001)

\bibitem{b2001}
A. Brandenburg, Astrophys. J., {\bf 550}, 824 (2001)

\bibitem{bf02} E.G. Blackman \& G.B. Field, 
Phys. Rev. Lett., {\bf 89}, 265007 (2002)

\bibitem{vishniac}  E. Vishniac \& J. Cho, ApJ, {\bf 550}, 752 (2001).

\bibitem{maronblackman}
J. Maron \& E.G. Blackman, Astrophys. J. Lett. {\bf 566}, L41 (2002). 

\bibitem{bb02} E.G. Blackman, 
\& A. Brandenburg  ApJ, 579, 359 (2002)

\bibitem{kleeorin02}N. Kleeorin, D. Moss, 
I. Rogachevskii, D. Sokoloff, D.Astron. Astrophys., {\bf 387}, 453 (2002)

\bibitem{bb03} E.G. Blackman 
\& A. Brandenburg, Astrophys. J. Lett., {\bf 584} L99  (2003)

\bibitem{blackman03}
E.G. Blackman, MNRAS, {\bf 344}, 707  (2003).

\bibitem{sb04} 
K. Subramanian, A. Brandenburg, Physical Review Letters, {\bf 93} 
205001  (2004)

\bibitem{bs05} 
A. Brandenburg, \& K. Subramanian, Phys. Rep., {\bf 417} 1 (2005)

\bibitem
{sur07a} S. Sur, K. Subramanian, 
\& A. Brandenburg, \mnras, {\bf 376}, 1238 (2007)

\bibitem{rad} K.-H. R\"adler, K.-H. Astron. Nachr., {\bf 301}, 101 (1980)

\bibitem{bf03} E.G. Blackman \& 
G.B. Field, Phys. Fluids, {\bf 15}, L73 (2003)

\bibitem{kr96}
N.I. Kleeorin,M.  Mond., \& I.V. Rogachevskii, Phys. Rev. E., {\bf 307}, 293
(1996) 

\bibitem{branfluid}
A. Brandenburg, P. K\"apyl\"a, A. Mohammed, 
 Phys. Fluids. {\bf 16}, 1020 (2004).

\bibitem{bf00b} E.G. Blackman,  \& G.B. Field, MNRAS, {\bf 318}, 
724 (2000)

\bibitem
{demoulin02} P. D{\'e}moulin, C.H. Mandrini, L. Van Driel-Gesztelyi,M.C. Lopez Fuentes \& G.Aulanier, Solar Physics, {\bf 207}, 87 (2002)

\bibitem{caramana84} E.J. Caramana, D.A. Baker,
Nuclear Fusion, {\bf 24}, 423 (1984)

\bibitem{japs94} H. Ji., A.F. Almagari, S.C. Prager, J.S. Sarff,
Phys. Rev. Lett. {\bf 73}, 668 (1994)

\bibitem{shukurov06} 
A. Shukurov, D. Sokoloff, K. Subramanian, A. Brandenburg, 
Astron. \& Astrophys. {\bf 448}, L33 (2006).

\bibitem
{sb06} 
K. Subramanian,  A. Brandenburg, ApJ {\bf 648}, L71 (2006)

\bibitem
{bsan} 
A. Brandenburg \& K. Subramanian,  Astronomische Nachrichten, {\bf 326}, 
400 (2005)

\bibitem
{b2005}
A. Brandenburg,  ApJ {\bf 625}, 539 (2005)

\bibitem{sur07b} S. Sur, A. Shukurov,  \& 
K. Subramanian,MNRAS in press (2007).

\bibitem{berger} M.A. Berger. \& G.B. Field  
J. Fluid. Mech, {\bf 147} 133 (1984).

\bibitem{finn} J.M. Finn \& T.M. Antonsen, Comments Plasma Phys. Controlled Fusion, {\bf 9}, 111123 (1985)


\end{document}